\documentclass[12pt]{article}
\usepackage{epsfig,a4}

 \advance\oddsidemargin by -1in
 \advance\evensidemargin
 by -1in
 \marginparsep
 0.4cm
 \advance\topmargin by
 -0.0in
 \makeindex

 \newcommand\la{\langle}
 \newcommand\ra{\rangle}
 \newcommand\beq{\begin{equation}}
 
 \newcommand\eeq{\end{equation}}                                               
 \newcommand\beqn{\begin{eqnarray}}
 \newcommand\eeqn{\end{eqnarray}}
 \newcommand\GeV{{\rm GeV}}

\def\BA{\begin{eqnarray}}
\def\BE{\begin{equation}}
\def\BF{\begin{figure}[htb]}
\def\BT{\begin{table}[htb]}
\def\EA{\end{eqnarray}}
\def\EE{\end{equation}}
\def\EF{\end{figure}}
\def\ET{\end{table}}

\def\la{\langle}
\def\ra{\rangle}
\def\mb{\,\mbox{mb}}
\def\fm{\,\mbox{fm}}
\def\GeV{\,\mbox{GeV}}

\def\lsim{\mathrel{\rlap{\lower4pt\hbox{\hskip1pt$\sim$}}
    \raise1pt\hbox{$<$}}}         
\def\gsim{\mathrel{\rlap{\lower4pt\hbox{\hskip1pt$\sim$}}
    \raise1pt\hbox{$>$}}}         

\begin{document} 
\vspace*{3cm}

\date{today}

\begin{center}
{\LARGE \bf
Study of Color Transparency 
\vspace*{0.2cm}
in Production  \\
\vspace*{0.2cm}
of Vector Mesons off Nuclei\footnote
{Based on the talk
presented at the International Conference HADRON STRUCTURE '2002,
Herl'any, Slovakia,
22-27 September, 2002.}} 
\end{center}


\begin{center}

\vspace{0.5cm}
 {\large J.~Nemchik}
 \\[1cm]
 {\sl Institute of Experimental Physics SAS, Watsonova 47,
04353 Kosice, Slovakia}

\end{center}

\vspace{1cm}
\begin{abstract}    

Within a light-cone QCD formalism 
based on the Green function
technique
incorporating color transparency, 
coherence length effects 
and gluon shadowing we study electroproduction
of vector mesons off nuclei.
We found rather large color transparency effects
in the range of $Q^2 \leq 10\div 20\GeV^2$.
They are stronger at low than at high energies 
and can be easily identified by 
HERMES or at JLab. 
We provide 
predictions for incoherent and coherent 
vector meson production for future
measurements.

\end{abstract}

\newpage



\section{Introduction} 
\label{intro}

One of the fundamental phenomenon coming from 
the quantum chromodynamics (QCD) is 
color transparency (CT),
which can be treated
either in the hadronic or in the quark basis.
The former approach leads to
Gribov's inelastic corrections \cite{gribov},
the latter one manifests itself as a result of 
color screening \cite{zkl,bbgg}. 
Although these two approaches are
complementary, we will use the quark-gluon interpretation,
which is more intuitive and straightforward. 

Investigation of diffractive electroproduction of vector mesons 
off nuclei is very effective for study of CT. 
A photon of high virtuality $Q^2$ is
expected to produce a pair with a small $\sim 1/Q^2$ transverse
separation\footnote{For production of light vector mesons
($\rho^0$, $\Phi^0$) very asymmetric pairs can be possible 
when either $q$ or $\bar q$ carries almost the whole
photon momentum.
As a results the $\bar qq$ pair can have a large separation,
see Sect.~\ref{lc} and Eq.~(\ref{212}). However, it is not so
for production of charmonia, where 
mainly symmetric $\bar qq$ pairs 
(either $q$ or $\bar q$ carries one half of the whole
photon momentum) dominate.}. 
Then CT manifests itself as 
a vanishing absorption of the
small sized colorless $\bar qq$ wave
packet during propagation through the nucleus. 
Dynamical evolution of the small sized $\bar qq$ pair to a normal
sized vector meson is controlled by the time scale, called formation
time.
Due to uncertainty principle, one needs 
a time interval to resolve
different levels $V$ (the ground state) or $V'$ 
(the next excited state) in the final state.
In the rest frame of the nucleus this formation time is
Lorentz dilated,
%
 \beq
t_f = \frac{2\,\nu}
{\left.m_{V^\prime}\right.^2 - m_V^2}\ ,
\label{20}
 \eeq
%
where $\nu$ is the photon energy.
A rigorous quantum-mechanical description of the pair evolution was
suggested in \cite{kz-91} and is based on the light-cone Green function
technique. 
A complementary description of the same process in the hadronic basis
is presented in \cite{hk-97}.

Another phenomenon known to cause nuclear
suppression is the effect of quantum coherence.
It results 
from destructive interference of
the amplitudes for which the interaction takes place on different bound
nucleons. 
Again, it can be estimated by relying on the uncertainty principle and
Lorentz time dilation as,
%
 \beq
t_c = \frac{2\,\nu}{Q^2 + m_V^2}\ .
\label{30}
 \eeq 
%
 It is usually called coherence time, but we also will use the term
coherence length (CL), since light-cone kinematics is assumed, $l_c=t_c$
(similarly, for formation length $l_f=t_f$). CL is related to the
longitudinal momentum transfer $q_c=1/l_c$ in $\gamma^*\,N \to V\,N$,
which controls the interference of the production amplitudes from
different nucleons.

At high energies (small Bjorken $x_{Bj}$) 
gluon shadowing becomes an important
phenomenon \cite{knst-01}. 
It was shown in \cite{knst-01,ikth-02} that for electroproduction of 
light vector mesons and charmonia off nuclei 
the gluon shadowing starts to be important at
center-of-mass system (c.m.s.) energies
$\sqrt{s} \geq 15\div 30\GeV$ and 
$\sqrt{s} \geq 30\div 60\GeV$, respectively,
depending on nuclear target and $Q^2$.
Although the gluon shadowing is quite small
in the kinematic range 
important for investigation of CT and discussed in the 
present paper we include it in all calculations.

In electroproduction of
vector mesons off nuclei one needs to disentangle CT (absorption) 
and CL (shadowing) as the two sources of nuclear suppression. 
The problem of CT-CL separation 
arises especially in production of light vector
mesons ($\rho^0$, $\Phi^0$) \cite{knst-01}.
In this case the coherence and formation lengths are
comparable 
starting from the photoproduction limit up to
$Q^2 \sim 1\div 2\GeV^2$.
In charmonium production, however, there is a strong
inequality $l_f > l_c$
independently of $Q^2$ and $\nu$ \cite{n-02}. It leads to
a different scenario of CT-CL mixing compared
to light vector meson production. 

Because a variation of
$l_c$ with $Q^2$ can mimic CT at medium and low energies, one can map
experimental events in $Q^2$ and $\nu$ in such a way as to keep
$l_c=const$. Then any rise of nuclear transparency,
$Tr_A = \sigma(\gamma^{*}A\to VX)/A~\sigma(\gamma^{*}N\to VN)$,
with $Q^2$ would give a signal for CT.
We will demonstrate that 
the LC dipole formalism based on the Green function
technique
predicts a large effect of CT in the range of $Q^2 \leq 10\div 20\GeV^2$
for both light vector meson \cite{knst-01}
and charmonium production \cite{n-02}.
This fact makes
it feasible to find a clear signal of CT effects also in exclusive 
production of vector mesons in the planned future experiments.

Calculations of coherent vector meson production off nuclei  
show that
the effect of CT on the $Q^2$
dependence of nuclear transparency at $l_c=const$ is weaker than in the
case of incoherent production and is difficult to be detected at low
energies since the cross section is small. 

%
\section{Light-cone dipole phenomenology for elastic 
electroproduction of vector meson \boldmath$\gamma^{*}N\to V~N$.}
\label{lc}
%

The light-cone (LC) dipole approach for elastic electroproduction 
$\gamma^{*}N\to V~N$ was used in 
\cite{hikt-00,n-02} to study the exclusive photo- and electroproduction
of charmonia and  
in \cite{knst-01} for elastic virtual photoproduction of
light vector mesons $\rho^0$ and $\Phi^0$.
Here a diffractive process
is treated as elastic scattering of a $\bar qq$ fluctuation
of the incident particle.  The elastic amplitude is given by 
convolution of the
universal flavor independent dipole cross section for the $\bar qq$
interaction with a nucleon, $\sigma_{\bar qq}$, \cite{zkl}
and the initial and final wave functions.  
It can be
represented in the quantum-mechanical form
%
 \BE
{\cal M}_{\gamma^{*}N\rightarrow V~N}(s,Q^{2}) =
\langle V |\sigma_{\bar qq}^N(\vec r,s)|\gamma^{*}\rangle=
\int\limits_{0}^{1} d\alpha \int d^{2}{{r}}\,\,
\Psi_{V}^{*}({\vec{r}},\alpha)\,   
\sigma_{\bar qq}({\vec{r}},s)\,  
\Psi_{\bar qq}({\vec{r}},\alpha,Q^2)\,
\label{120}
 \EE
%
 with the normalization 
%
 \beq
\left.\frac{d\sigma}{dt}\right|_{t=0} =
\frac{|{\cal M}|^{2}}{16\,\pi}.
\label{125}
 \eeq
%

In order to calculate the photoproduction amplitude one needs to know the
following ingredients of Eq.~(\ref{120}): 
(i)  
the dipole cross section
$\sigma_{\bar qq}({\vec{r}},s)$ which depends on the $\bar qq$
transverse separation $\vec{r}$ and the c.m. energy squared $s$. 
(ii)  
The light-cone (LC)  wave function of the $\bar qq$ Fock component of 
the photon
$\Psi_{\bar qq}({\vec{r}},\alpha,Q^2)$ which also depends on the
photon virtuality $Q^2$ and the relative share $\alpha$ of the photon
momentum carried by the quark. 
(iii) 
The LC vector meson wave function
$\Psi_{V}(\vec r,\alpha)$. 

Note that in the LC formalism the photon and meson wave functions contain
also higher Fock states $|\bar qq\ra$, $|\bar qqG\ra$, $|\bar qq2G\ra$,
etc. 
The effects of higher Fock states are implicitly incorporated
into the energy dependence of the dipole cross section
$\sigma_{\bar qq}(\vec r,s)$ as is given in Eq.~(\ref{120}).

The dipole cross section $\sigma_{\bar qq}(\vec r,s)$ 
represents the interaction of a
$\bar qq$ dipole of transverse separation $\vec r$ with a nucleon
\cite{zkl}.
It is a flavor independent universal function of
$\vec{r}$ and energy and allows to describe 
various high energy processes
in an uniform way.
It is known to vanish quadratically
$\sigma_{\bar qq}(r,s)\propto r^2$ as $r\rightarrow 0$ due to color
screening (CT property) and cannot be predicted
reliably because of poorly known higher order 
perturbative QCD (pQCD) corrections and
nonperturbative effects. 
Detailed discussion about the dipole cross section 
$\sigma_{\bar qq}(\vec r,s)$ 
with an emphasis on the production of light vector mesons
is presented in \cite{knst-01}.
In electroproduction of charmonia \cite{n-02} the corresponding 
transverse
separations of $\bar cc$-dipole reach the values $\leq 0.4\fm$
(semiperturbative region). It means that nonperturbative effects
are sufficiently smaller as compared with light vector mesons.
Similarly, the relativistic corrections are also small enough 
and the nonrelativistic limit $\alpha = 0.5$ can be safely
used with rather high accuracy \cite{kz-91}.

There are two popular parameterizations of $\sigma_{\bar qq}(\vec 
r,s)$, GBW presented in \cite{gbw} and KST suggested in \cite{kst2}.
We choose the second parametrization, because it is 
valid down to the limit of real photoproduction.

The dipole cross section $\sigma_{\bar qq}(\vec r,s)$ provides the
imaginary part of the elastic amplitude.  It is known, however, that the
energy dependence of the total cross section generates also a real part
\cite{bronzan},
%
 \beq
\sigma_{\bar qq}(r,s) \Rightarrow
\left(1-i\,\frac{\pi}{2}\,\frac{\partial}
{\partial\ln(s)}\right)\,
\sigma_{\bar qq}(r,s).
\label{real-part}
 \eeq
%

The perturbative distribution amplitude (``wave function'') of the $\bar
qq$ Fock component of the photon 
has the following form 
for transversely (T) and longitudinally (L) polarized photons 
\cite{lc,bks-71,nz-91},
%
 \BE
\Psi_{\bar qq}^{T,L}({\vec{r}},\alpha) =
\frac{\sqrt{N_{C}\,\alpha_{em}}}{2\,\pi}\,\,
Z_{q}\,\bar{\chi}\,\hat{O}^{T,L}\,\chi\, 
K_{0}(\epsilon\,r)
\label{70}
 \EE
%
 where $\chi$ and $\bar{\chi}$ are the spinors of the quark and
antiquark, respectively; $Z_{q}$ is the quark charge,
$N_{C} = 3$ is the
number of colors. $K_{0}(\epsilon r)$ is a modified Bessel
function with 
%
 \BE
\epsilon^{2} =
\alpha\,(1-\alpha)\,Q^{2} + m_{q}^{2}\ ,
\label{80}
 \EE
%
where $m_{q}$ is the quark mass  
and $\alpha$ is the 
fraction of the LC momentum of the photon carried by the quark. The operators
$\widehat{O}^{T,L}$ read,
%
 \BE 
\widehat{O}^{T} = m_{q}\,\,\vec{\sigma}\cdot\vec{e} +
i\,(1-2\alpha)\,(\vec{\sigma}\cdot\vec{n})\,
(\vec{e}\cdot\vec{\nabla}_r) + (\vec{\sigma}\times
\vec{e})\cdot\vec{\nabla}_r\ ,
 \label{90}
 \EE
%
%
 \BE
\widehat{O}^{L} =
2\,Q\,\alpha (1 - \alpha)\,(\vec{\sigma}\cdot\vec{n})\ .
\label{100}
 \EE
%
 Here $\vec\nabla_r$ acts on transverse coordinate $\vec r$;
$\vec{e}$ is the polarization vector of the photon, $\vec{n}$ is a unit
vector parallel to the photon momentum and $\vec{\sigma}$ is the three
vector of the Pauli spin-matrices.

In general,
the transverse $\bar qq$ separation is controlled by the distribution 
amplitude Eq.~(\ref{70}) with the mean value,
%
 \BE
\la r\ra \sim \frac{1}{\epsilon} = 
\frac{1}{\sqrt{Q^{2}\,\alpha\,(1-\alpha) + m_{q}^{2}}}\,.
\label{212}
 \EE
%

For production of light vector mesons
very asymmetric $\bar qq$ pairs with $\alpha$ or $(1-\alpha) \lsim
m_q^2/Q^2$ become possible. Consequently,
the mean transverse separation $\la r\ra \sim 1/m_q$ becomes
huge since one must use current quark masses within pQCD.
A popular recipe how to fix this problem is to introduce an effective
quark mass $m_{eff}\sim\Lambda_{QCD}$ which should represent
the nonperturbative interaction effects between $q$ and $\bar q$.
However, we introduce this interaction explicitly taking 
from \cite{kst2} the corresponding phenomenology 
based on the light-cone Green function approach.
 
The Green function $G_{\bar qq}(z_1,\vec r_1;z_2,\vec r_2)$ 
describes the 
propagation of an interacting $\bar qq$ pair 
($\bar cc$ pair for the case of $J/\Psi$ production)
between points with
longitudinal coordinates $z_{1}$ and $z_{2}$ and with initial and final
separations $\vec r_1$ and $\vec r_2$. This
Green function satisfies the 
two-dimensional Schr\"odinger equation, 
%
 \BE
i\frac{d}{dz_2}\,G_{\bar qq}(z_1,\vec r_1;z_2,\vec r_2)=
\left[\frac{\epsilon^{2} - \Delta_{r_{2}}}{2\,\nu\,\alpha\,(1-\alpha)}
+V_{\bar qq}(z_2,\vec r_2,\alpha)\right]
G_{\bar qq}(z_1,\vec r_1;z_2,\vec r_2)\ .
\label{250}
 \EE
%
 Here $\nu$ is the photon energy. The Laplacian $\Delta_{r}$ acts on
the coordinate $r$.  

The imaginary part of the LC potential $V_{\bar
qq}(z_2,\vec r_2,\alpha)$ in (\ref{250}) is responsible for
attenuation of the $\bar qq$ in the medium, while the real part
represents the interaction between the $q$ and $\bar{q}$.  
This potential is supposed to provide the correct LC wave functions of 
vector mesons. For the sake of simplicity we use  the oscillator form 
of the potential,
%
 \BE  
{\rm Re}\,V_{\bar qq}(z_2,\vec r_{2},\alpha) =
\frac{a^4(\alpha)\,\vec r_{2}\,^2} 
{2\,\nu\,\alpha(1-\alpha)}\ ,
\label{260}
 \EE
%
 which leads to a Gaussian $r$-dependence of the LC wave function of the
meson ground state.  The shape of the function $a(\alpha)$ is described
in \cite{kst2} and discussed in \cite{knst-01}
for vector meson production.

 In this case equation (\ref{250}) has an analytical solution, the
harmonic oscillator Green function \cite{fg},
%
 \BA 
G_{\bar qq}(z_1,\vec r_1;z_2,\vec r_2) &=&
\frac{a^2(\alpha)}{2\,\pi\,i\,
{\rm sin}(\omega\,\Delta z)}\, {\rm exp}
\left\{\frac{i\,a^2(\alpha)}{{\rm sin}(\omega\,\Delta z)}\,
\Bigl[(r_1^2+r_2^2)\,{\rm cos}(\omega \,\Delta z) -
2\,\vec r_1\cdot\vec r_2\Bigr]\right\}
\nonumber\\ \times \,
& &{\rm exp}\left[- 
\frac{i\,\epsilon^{2}\,\Delta z}
{2\,\nu\,\alpha\,(1-\alpha)}\right] \ , 
\label{270} 
 \EA
%
%
where $\Delta z=z_2-z_1$ and 
%
 \BE 
\omega = \frac{a^2(\alpha)}{\nu\;\alpha(1-\alpha)}\ .
\label{280}
 \EE
%
 The boundary condition is $G_{\bar
qq}(z_1,\vec r_1;z_2,\vec r_2)|_{z_2=z_1}=
\delta^2(\vec r_1-\vec r_2)$.

The probability amplitude to find the $\bar qq$ fluctuation of a photon
at the point $z_2$ with separation $\vec r$ is given by an integral
over the point $z_1$ where the $\bar qq$ is created by the photon with
initial zero separation,
%
 \BE
\Psi^{T,L}_{\bar qq}(\vec r,\alpha)=
\frac{i\,Z_q\sqrt{\alpha_{em}}}
{4\pi\,\nu\,\alpha(1-\alpha)} 
\int\limits_{-\infty}^{z_2}dz_1\,
\Bigl(\bar\chi\;\widehat O^{T,L}\chi\Bigr)\,
G_{\bar qq}(z_1,\vec r_1;z_2,\vec r)
\Bigr|_{r_1=0}\ .
\label{290}
 \EE
%
 The operators $\widehat O^{T,L}$ are defined in Eqs.~(\ref{90}) and
(\ref{100}). Here they act on the coordinate $\vec r_1$.

If we write the transverse part  as
%
 \BE
\bar\chi\;\widehat O^{T}\chi= 
\bar\chi\;m_{q}\,\,\vec{\sigma}\cdot\vec{e}\,\chi +
\bar\chi\;\left[i\,(1-2\alpha)\,(\vec{\sigma}\cdot\vec{n})\,
\vec{e} + (\vec{\sigma}\times
\vec{e})\right]\,\chi\cdot\vec{\nabla}_{r}=
E+\vec F\cdot\vec\nabla_{r}\ ,
\label{300}
 \EE
%
 then the distribution functions read,   
%
 \BE
\Psi^{T}_{\bar qq}(\vec r,\alpha) =
Z_q\sqrt{\alpha_{em}}\,\left[E\,\Phi_0(\epsilon,r,\lambda)
+ \vec F\,\vec\Phi_1(\epsilon,r,\lambda)\right]\ ,
\label{310}
 \EE
%
%
 \BE
\Psi^{L}_{\bar qq}(\vec r,\alpha) =
2\,Z_q\sqrt{\alpha_{em}}\,Q\,\alpha(1-\alpha)\,
\bar\chi\;\vec\sigma\cdot\vec n\;\chi\,
\Phi_0(\epsilon,r,\lambda)\ ,
\label{320}
 \EE
%
 where
%
 \BE
\lambda=
\frac{2\,a^2(\alpha)}{\epsilon^2}\ .
\label{330}
 \EE
%

The functions $\Phi_{0,1}$ in Eqs.~(\ref{310}) and (\ref{320})
are defined as
%
 \BE
\Phi_0(\epsilon,r,\lambda) =
\frac{1}{4\pi}\int\limits_{0}^{\infty}dt\,
\frac{\lambda}{{\rm sh}(\lambda t)}\,
{\rm exp}\left[-\ \frac{\lambda\epsilon^2 r^2}{4}\,
{\rm cth}(\lambda t) - t\right]\ ,
\label{340}
 \EE
%
%
 \BE
\vec\Phi_1(\epsilon,r,\lambda) =
\frac{\epsilon^2\vec r}{8\pi}\int\limits_{0}^{\infty}dt\,
\left[\frac{\lambda}{{\rm sh}(\lambda t)}\right]^2\,
{\rm exp}\left[-\ \frac{\lambda\epsilon^2 r^2}{4}\,
{\rm cth}(\lambda t) - t\right]\ .  
\label{350}
 \EE
%

Note that the $\bar q-q$ interaction enters Eqs.~(\ref{310}) and
(\ref{320}) via the parameter $\lambda$ defined in (\ref{330}). In the
limit of vanishing interaction $\lambda\to 0$ (i.e. $Q^2\to \infty$,
$\alpha$ is fixed, $\alpha\not=0$ or $1$) Eqs.~(\ref{310}) -
(\ref{320})  produce the perturbative expressions of Eq.~(\ref{70}).
For charmonium production 
nonperturbative interaction effects are quite weak. Consequently, 
the parameter $\lambda$ (\ref{330}) 
is rather small due to a large mass of the $c$ quark.

The last ingredient in elastic production amplitude 
(\ref{120}) is the vector meson wave function.
We use a popular prescription
\cite{terentev} applying the Lorentz boost to the rest frame wave
function assumed to be Gaussian which leads to radial parts of
transversely and longitudinally polarized mesons in the form,
%
 \BE
\Phi_{V}^{T,L}(\vec r,\alpha) = 
C_{V}^{T,L}\,\alpha(1-\alpha)\,f(\alpha)\,
{\rm exp}\left[-\ \frac{\alpha(1-\alpha)\,{r}^2}
{2\,R^2}\right]\  
\label{170}
 \EE
%
 with a normalization defined below, and 
%
 \beq
f(\alpha) = \exp\left[-\ \frac{m_q^2\,R^2}
{2\,\alpha(1-\alpha)}\right]\ 
\label{170a}
 \eeq 
%
with the parameters from \cite{jan97}, $R=0.183\,\fm$,
$m_q=m_c=1.5\GeV$ for $J/\Psi$ and
$R=0.590\,\fm$,   
$m_q=0.150\GeV$ for $\rho^0$.

We assume that the distribution amplitude of $\bar qq$ fluctuations for
vector meson and for the photon have a similar structure
\cite{jan97}.  Then in analogy to Eqs.~(\ref{310}) -- (\ref{320}),
%
 \beqn
\Psi^T_{V}(\vec r,\alpha) &=&
(E+\vec F\cdot\vec\nabla_{r})\,
\Phi^T_{V}(\vec r,\alpha)\ ;
\label{172}\\
\Psi^L_{V}(\vec r,\alpha) &=& 
2\,m_{V}\,\alpha(1-\alpha)\,
(\bar\chi\,\vec\sigma\cdot\vec n\,\chi)\,
\Phi^L_{V}(\vec r,\alpha)\ .
\label{174}
 \eeqn
%

Correspondingly, the normalization conditions for the transverse 
and longitudinal vector meson wave
functions read,
%
 \BE
N_{C}\,\int d^{2} r\,\int d\alpha \left\{ m_{q}^{2}\,
\Bigl|\Phi^T_{V}(\vec r,\alpha)\Bigr|^{2} + \Bigl[\alpha^{2} + 
(1-\alpha)^{2}\Bigr]\,
\Bigl|\partial_{r}
\Phi^T_{V}(\vec r,\alpha)\Bigr|^{2} \right\} = 1
\label{230}
 \EE
%
%
 \BE
4\,N_{C}\,\int d^{2} r\,\int d\alpha\, 
\alpha^{2}\,(1-\alpha)^{2}\,
m_{V}^{2}\,\Bigl|\Phi^L_{V}(\vec r,\alpha)\Bigr|^2 = 1\, .
\label{240}
 \EE
%

%
\section{Electroproduction of vector mesons on a nucleon}
\label{data-N}
%

The forward production amplitude
$\gamma^*\,N \to V\,N$ for transverse and longitudinal photons and 
vector mesons is calculated
using the nonperturbative photon Eqs.~(\ref{310}),
(\ref{320}) and vector meson Eqs.~(\ref{172}), (\ref{174}) wave functions  
and has the following form,
%
 \BA
{\cal M}_{\gamma^{*}N\rightarrow V\,N}^{T}(s,Q^{2})
\Bigr|_{t=0} &=& 
N_{C}\,Z_{q}\,\sqrt{2~\alpha_{em}}
\int d^{2} r\,\sigma_{\bar qq}(\vec r,s)
\int\limits_0^1 d\alpha \Bigl\{ m_{q}^{2}\,
\Phi_{0}(\epsilon,\vec r,\lambda)\Phi^T_{V}(\vec r,\alpha)
\nonumber\\ 
&-& \bigl [\alpha^{2} + (1-\alpha)^{2}\bigr ]\,
\vec{\Phi}_{1}(\epsilon,\vec r,\lambda)\cdot
\vec{\nabla}_{r}\,\Phi^T_{V}(\vec r,\alpha) \Bigr\}\,;
\label{360}
 \EA
%
%
 \BA
{\cal M}_{\gamma^{*}N\rightarrow V\,N}^{L}(s,Q^{2})
\Bigr|_{t=0} &=& 
4\,N_{C}\,Z_{q}\,\sqrt{2~\alpha_{em}}\,m_{V}\,Q\,
\int d^{2} r\,\sigma_{\bar qq}(\vec r,s)
\nonumber\\&\times& 
\int\limits_0^1 d\alpha\,
\alpha^{2}\,(1-\alpha)^{2}\, 
\Phi_{0}(\epsilon,\vec r,\lambda)
\Phi^L_{V}(\vec r,\alpha)\ .
\label{370}
 \EA
%
 These amplitudes are normalized as ${|{\cal M}^{T,L}|^{2}}=
\left.16\pi\,{d\sigma_{N}^{T,L}/ dt}\right|_{t=0}$. The real
part of the amplitude is included 
according to the prescription described in the previous section.
We calculate the cross sections
$\sigma = \sigma^T + \epsilon'\,\sigma^L$ assuming that the photon
polarization is $\epsilon'=1$.

We checked in \cite{knst-01} and \cite{n-02} the absolute 
value of the production cross section by
comparing with data for elastic electroproduction of 
light vector mesons and charmonia.
We found a good agreement of the model predictions with the data
on the $Q^2$ dependence of the cross section. 
As the second test of
our approach we compared also the model calculations with the data on
the real vector meson photoproduction \cite{knst-01,n-02}.
Successful description of those data confirms an importance
(especially for light vector mesons)
of the nonperturbative interaction effects 
between the $q$ and $\bar q$ 
included into calculations.

As a cross-check for the choice of the vector meson wave function in 
Eqs.~(\ref{170}) and (\ref{170a}) we also calculated the total
$V$-nucleon cross section, which has the following form,
%
 \BA
\sigma_{tot}^{V\,N} = 
N_{C}\int d^{2} r
\int d\alpha \left\{ m_{q}^{2}\,
\Bigl |\Phi^T_{V} (\vec r,\alpha)\Bigr |^2 +
\bigl [\alpha^{2} + (1-\alpha)^{2}\bigr ]\,
\Bigl |\partial_{r}\Phi^T_{V}(\vec r,\alpha)\Bigr|^{2} 
\right\}
\sigma_{\bar qq}(\vec r,s)
\label{380}
 \EA
%
At $\sqrt{s} = 10\GeV$ we obtain \cite{n-02}
$\sigma_{tot}^{J/\Psi\,N} =
4.2\,\mb$ which is not in contradiction with 
$\sigma_{tot}^{J/\Psi\,N} = 3.6\pm 0.1\,\mb$ evaluated in 
\cite{hikt-00}
using more realistic $\bar q-q$ potentials and/or
charmonium wave functions.
For $\rho^0$ we obtained \cite{knst-01}
$\sigma_{tot}^{\rho\,N} =
26.5\,\mb$, which is very close to pion-nucleon total cross section. 

%
\section{Incoherent production of vector mesons off nuclei}
\label{psi-incoh}
%

In diffractive incoherent (quasielastic) production of vector mesons
off nuclei, $\gamma^{*}\,A\rightarrow V\,X$, 
one sums over all final states of the target nucleus except those which 
contain particle (pion) creation. 
The observable usually studied experimentally is nuclear transparency 
defined as
%
 \BE
Tr^{inc}_{A} = 
\frac{\sigma_{\gamma^{*}A\to VX}^{inc}}
{A\,\sigma_{\gamma^{*}N\to VN}}\ .
\label{480}
 \EE
%
 The $t$-slope of the differential quasielastic cross section is the same
as on a nucleon target. Therefore, instead of integrated cross sections
one can also use nuclear transparency expressed via
the forward differential cross sections 
Eq.~(\ref{125}),
%
 \beq 
Tr^{inc}_A = \frac{1}{A}\,
\left|\frac{{\cal M}_{\gamma^{*}A\to VX}(s,Q^{2})}
{{\cal M}_{\gamma^{*}N\to VN}(s,Q^{2})}\right|^2\, .
\label{485}
 \eeq 
%

In the LC Green function approach \cite{knst-01}
the physical photon 
$|\gamma^*\ra$ is decomposed 
into different Fock states, namely, the bare photon
$|\gamma^*\ra_0$, $|\bar qq\ra$, $|\bar qqG\ra$, etc. 
As we mentioned above the higher Fock states
containing gluons describe the energy dependence of the
photoproduction reaction on a nucleon. 
Besides, those Fock components also lead to gluon shadowing  
as far as nuclear effects are concerned.
However, these fluctuations are heavier and have a
shorter coherence time (lifetime) than the lowest $|\bar qq\ra$
state. Therefore, at medium energies only $|\bar qq\ra$ fluctuations of 
the photon matter. Consequently, gluon shadowing related to the higher Fock
states will be dominated at high energies.
Detailed description and calculation of gluon shadowing 
for the case of vector meson production off nuclei is presented
in \cite{knst-01,ikth-02}.
Although the gluon shadowing effects are rather small 
in the kinematic range important 
for study of CT effects 
we include them in all calculations.

Propagation of an interacting $\bar qq$ pair in a nuclear medium is also
described by the Green function satisfying the evolution Eq.~(\ref{250}).
However, the potential in this case acquires an imaginary part which
represents absorption in the medium,
%
 \BE
Im V_{\bar qq}(z_2,\vec r,\alpha) = - 
\frac{\sigma_{\bar qq}(\vec r,s)}{2}\,\rho_{A}({b},z_2)\, ,
\label{440}
 \EE
%
where $\rho_{A}({b},z_2)$ is the nuclear density function defined
at the point with longitudinal coordinate $z_2$ and impact
parameter $\vec{b}$. 

The analytical solution of Eq.~(\ref{270}) is only known for the 
harmonic oscillator potential $V(r)\propto r^2$. To keep
the calculations reasonably simple we are forced to use the dipole 
approximation
%
 \beq
\sigma_{\bar qq}(r,s) = C(s)\,r^2\ ,
\label{460}
 \eeq
%
which allows to obtain the Green function in an analytical form.
The energy dependent factor $C(s)$ in Eq.~(\ref{460}) is adjusted 
by the procedure described in \cite{knst-01}.

With the potential Eqs.~(\ref{440}) -- (\ref{460}) the solution of 
Eq.~(\ref{250}) has the same form as Eq.~(\ref{270}), except that one 
should 
replace $\omega \Rightarrow \Omega$ 
and $a^2(\alpha) \Rightarrow b(\alpha)$, where
%
 \beq
\Omega = \frac{b(\alpha)}{\nu\;\alpha(1-\alpha)} =
\frac{\sqrt{a^4(\alpha)-
i\,\rho_{A}({b},z)\,
\nu\,\alpha\,(1-\alpha)\,C(s)}}
{\nu\;\alpha(1-\alpha)} \ .
\label{470}
 \eeq
%

As we discussed in \cite{knst-01} the value of $l_c$ can 
distinguish different regimes of vector meson production.

{\bf (i)} The CL is much shorter than the mean nucleon spacing in a
nucleus ($l_c \to 0$). 
For light vector mesons $l_f\sim l_c$ and 
consequently $l_f \to 0$ as well. 
In this case $G(z_2,\vec r_2;z_1,\vec r_1)  \to
\delta(z_2-z_1)$. 
$Tr_{A}^{inc}$ is given by the
simple formula corresponding to the Glauber
approximation \cite{knst-01,n-02}.

{\bf (ii)}
In production of charmonia and other heavy flavor quarkonii
there is a strong inequality $l_c < l_f$ and
the intermediate case $l_c\to 0$, but $l_f\sim R_A$ 
($R_A$ is the nuclear radius)
can be realized. 
Then the formation of the meson wave function 
is described by the Green function and the numerator of the nuclear 
transparency ratio Eq.~(\ref{485}) has the form 
\cite{kz-91},
%
 \beq
\Bigl|{\cal M}_{\gamma^{*}A\to VX}(s,Q^{2})
\Bigr|^2_{l_c\to0;\,l_f\sim R_A} = 
\int d^2b\int_{-\infty}^{\infty} dz\,\rho_A(b,z)\,
\Bigl|F_1(b,z)\Bigr|^2\ ,
\label{500}
 \eeq
%
 where
%
 \beq
F_1(b,z) = 
\int_0^1 d\alpha
\int d^{2} r_{1}\,d^{2} r_{2}\,
\Psi^{*}_{V}(\vec r_{2},\alpha)\,
G(z^\prime,\vec r_{2};z,\vec r_{1})\,
\sigma_{\bar qq}(r_{1},s)\,
\Psi_{\bar qq}(\vec r_{1},
\alpha)\Bigl|_{z^\prime\to\infty}
\label{505}
 \eeq

{\bf (iii)} 
In the high energy limit, the CL
$l_c \gg R_A$.
In this case $G(z_2,\vec r_2;z_1,\vec r_1)
\to \delta(\vec r_2 - \vec r_1)$, i.e. all fluctuations of 
the transverse $\bar qq$ 
separation are ``frozen'' by Lorentz time dilation.
Then, the numerator on the r.h.s. of Eq.~(\ref{485}) takes the form
\cite{kz-91},
%
 \beqn
\Bigl|{\cal M}_{\gamma^{*}A\to VX}(s,Q^{2})
\Bigr|^2_{l_c \gg R_A} = 
\int d^2b\,T_A(b)\left|\int d^2r\int_0^1 d\alpha 
\right. 
\label{510}\\
\times \left.\Psi_{V}^{*}(\vec r,\alpha)\,
\sigma_{\bar qq}(r,s)\,  
\exp\left[-{1\over2}\sigma_{\bar qq}(r,s)\,T_A(b)\right]
\Psi_{\bar qq}(\vec r,\alpha,Q^2)\right|^2\ ,
\nonumber
 \eeqn 
%
where $T_A(b) = \int_{-\infty}^{\infty} dz\,\rho_A(b,z)$
is the nuclear thickness function.

{\bf (iv)} 
This regime reflects the general case when there is
no restrictions for either $l_c$ or $l_f$.  
The corresponding theoretical tool 
has been developed for the first time in \cite{knst-01}.
In this general case
the incoherent photoproduction amplitude is
represented as a sum of two terms \cite{hkz},
%
 \BE 
\Bigl|\,{\cal M}_{\gamma^{*}A\to
VX}(s,Q^{2})\Bigr|^{2} = \int d^{2}b
\int\limits_{-\infty}^{\infty} dz\,\rho_{A}({b},z)\, 
\Bigl|F_{1}({b},z) - F_{2}({b},z)\Bigr|^{2}\ .
\label{520}
 \EE
%
 The first term $F_{1}({b},z)$ introduced above in Eq.~(\ref{505}) 
alone would correspond to the short
$l_c$ limit (ii). The second term $F_{2}({b},z)$ in Eq.~(\ref{520})  
corresponds to the situation when the
incident photon produces a $\bar qq$ pair diffractively and coherently at
the point $z_1$ prior to incoherent quasielastic scattering at point $z$.
The LC Green functions describe the evolution of the $\bar qq$ over the
distance from $z_1$ to $z$ and further on, up to the formation of the
meson wave function. Correspondingly, this term has the form,
%
 \beqn
F_{2}(b,z) &=& \frac{1}{2}\,
\int\limits_{-\infty}^{z} dz_{1}\,\rho_{A}(b,z_1)\,
\int\limits_0^1 d\alpha\int d^2 r_1\,
d^2 r_{2}\,d^2 r\,
\Psi^*_V (\vec r_2,\alpha)
\nonumber \\
&\times&
G(z^{\prime}\to\infty,\vec r_2;z,\vec r)\,
\sigma_{\bar qq}(\vec r,s)\,
G(z,\vec r;z_1,\vec r_1)\,
\sigma_{\bar qq}(\vec r_1,s)\,
\Psi_{\bar qq}(\vec r_1,\alpha)\, .
\label{530}
 \eeqn
%
Eq.~(\ref{520}) correctly reproduces the limits (i) - (iii). 
\vspace*{0.3cm}

Exclusive incoherent electroproduction of vector mesons off nuclei has
been suggested in \cite{knnz} for investigation of CT. 
Increasing
the photon virtuality $Q^2$ one squeezes the produced $\bar qq$ wave
packet. Such a small colorless system propagates through the nucleus with
little attenuation, provided that the energy is sufficiently high
($l_f\gg R_A$) so the fluctuations of the $\bar qq$ separation are frozen 
during propagation.
Consequently, a rise of nuclear transparency
$Tr_A^{inc}(Q^2)$ with $Q^2$ should give a signal for CT.
Indeed, such a rise was observed in the E665 experiment \cite{e665-rho}
at Fermilab for exclusive production of $\rho^0$ mesons off
nuclei what has been claimed as manifestation of CT.
%
 \begin{figure}[tbh]
\includegraphics{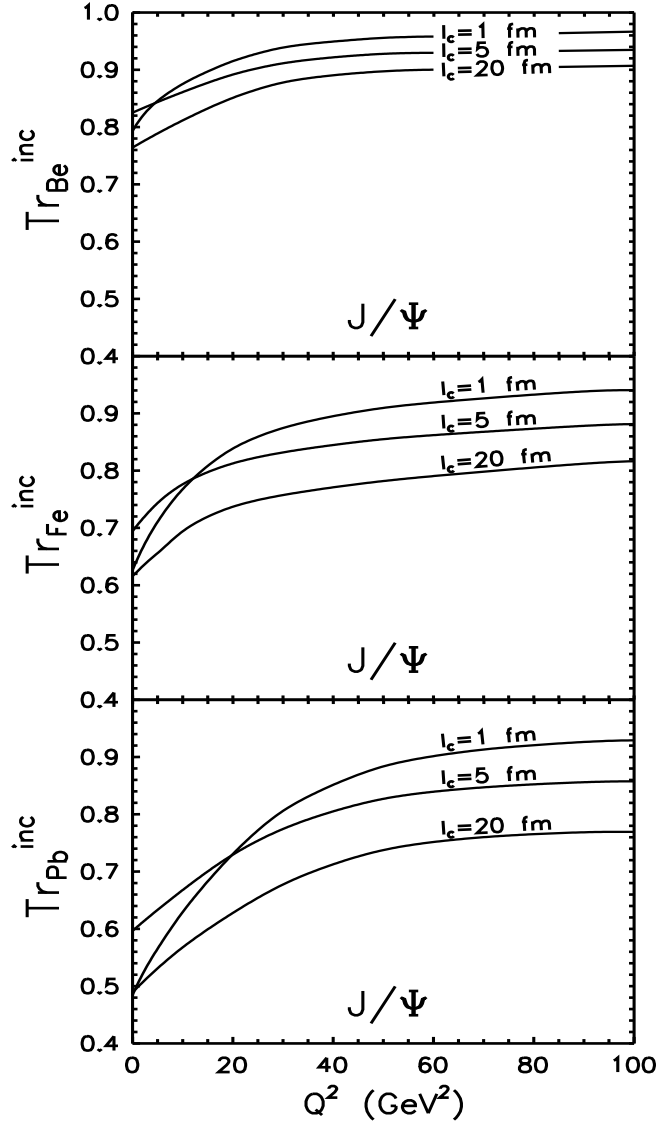}
\begin{center}
\vspace{14.7cm}
\parbox{13cm}
{\caption[Delta]
 {$Q^2$ dependence of the nuclear transparency $Tr_A^{inc}$ for
exclusive electroproduction of $J/\Psi$
on nuclear targets $^{9}Be$, $^{56}Fe$ and $^{207}Pb$ (from top to
bottom). The CL is fixed at $l_c = 1$, $5$ and
$20\fm$.}
 \label{lc-const-inc}}
\end{center}
 \end{figure}
%

However, the effect of coherence length \cite{kn95,hkn} leads also
to a rise of $Tr_A^{inc}(Q^2)$ with $Q^2$ and so can imitate
CT effects.
This happens when the coherence length varies from long to short 
(see Eq.~(\ref{30}))
compared to the nuclear size and  
the length of the path in nuclear matter becomes shorter.
Consequently, the vector meson (or $\bar qq$)
attenuates less in nuclear medium.
This happens when $Q^2$ increases at fixed $\nu$.
Therefore one should carefully disentangle these two phenomena.

%
 \begin{figure}[tbh]
\includegraphics{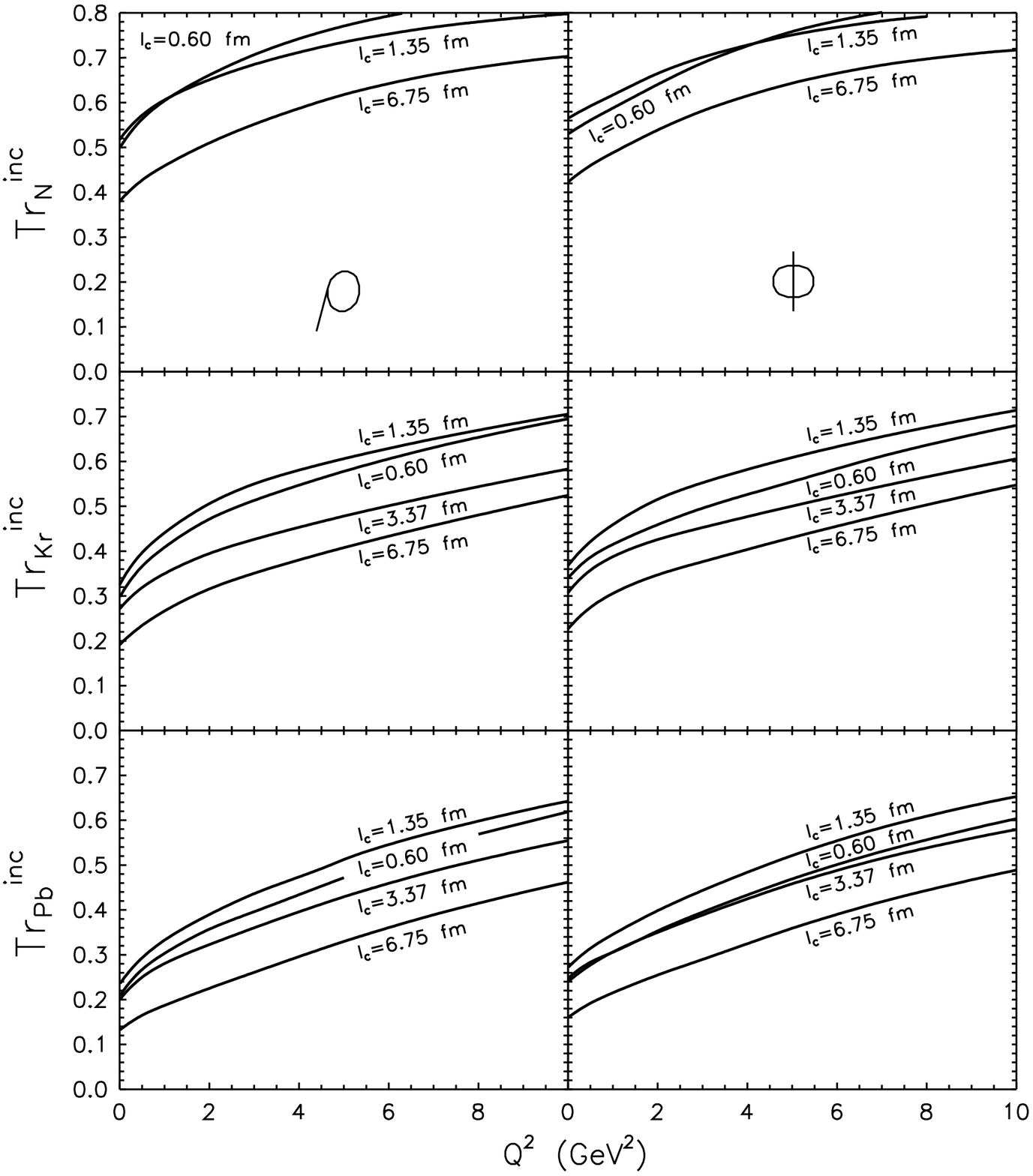}
\begin{center}
\vspace{15.3cm}
\parbox{13cm}
{\caption[Delta]
 {$Q^2$ dependence of the nuclear transparency $Tr_A^{inc}$ for
exclusive electroproduction of $\rho$ (left) and $\Phi$ (right)
mesons
on nuclear targets $^{14}N$, $^{84}Kr$ and $^{207}Pb$ (from top to
bottom). The CL is fixed at $l_c = 0.60$, $1.35$, $3.37$ and
$6.75\fm$.}
 \label{lc-const-inc-all}}
\end{center}
 \end{figure}
%

Model calculations of incoherent $\rho$ and $J/\Psi$ production
off nuclei were tested in comparison with available data
and a nice agreement has been found \cite{knst-01,n-02}. 
However, the available data do not solve the
problem of separation of CT and CL effects
discussed in \cite{knst-01}.
Because of
$l_c \gsim l_f$ at $Q^2 \lsim 1\div 2\GeV^2$
for production of $\rho^0$ and $\Phi^0$ \cite{knst-01}
and
a strong inequality $l_c < l_f$
for charmonium production \cite{n-02},
there is a different scenario of CT-CL mixing for
light and heavy vector mesons production.

In order to eliminate
the effect of CL from the data
on the $Q^2$ dependence of nuclear transparency
one should simply bin the data in a
such way which keeps $l_c = const$ \cite{hk-97}.
It means that one should vary simultaneously $\nu$
and $Q^2$ maintaining the CL Eq.~(\ref{30}) constant,
%
 \beq
\nu = {1\over2}\,l_c\,(Q^2+m_{V}^2)\ .
\label{534}
 \eeq
%
 In this case the Glauber model predicts a $Q^2$ independent nuclear
transparency, and any rise with $Q^2$ would signal CT \cite{hk-97}.

The LC Green function technique incorporates both the effects of
coherence and formation. We performed calculations of $Tr_A^{inc}(Q^2)$
at fixed $l_c$ starting from different minimal values of $\nu$, which
correspond to real photoproduction in Eq.~(\ref{534}),
%
 \beq
\nu_{min}={1\over2}\,l_c\,m_{V}^2\ . 
\label{536}
 \eeq
%
The results for incoherent production of $J/\Psi$ at 
$\nu_{min}= 24.3,\ 121.7$ and $487\GeV$ ($l_c=1, 5$ and 
$20\fm$) are presented in Fig.~\ref{lc-const-inc} for beryllium, 
iron and lead. 
We use the nonperturbative LC wave function of the photon with the
parameters of the LC potential described in \cite{n-02}.
We use quark mass $m_q=m_c=1.5\GeV$. 

The analogical results for incoherent production of $\rho$
and $\Phi$ mesons at $l_c=0.60$, $1.35$, $3.37$ and 
$6.75\fm$ are presented in Fig.~\ref{lc-const-inc-all} for nitrogen, 
krypton and lead. 

The both Figs. ~\ref{lc-const-inc} and
~\ref{lc-const-inc-all} exhibit large CT effects.
Although the predicted $Q^2$- variation of nuclear transparency 
at fixed $l_c$ for $J/\Psi$ 
is less than for light vector
mesons, it is still sufficiently significant
to be investigated experimentally even in the range of $Q^2\lsim
20\GeV^2$. 
CT effects (the rise with $Q^2$ of nuclear transparency)
are more pronounced at low than at high energies
and can be easily identified by the planned future
experiments.
%
 \begin{figure}[htb]
\includegraphics{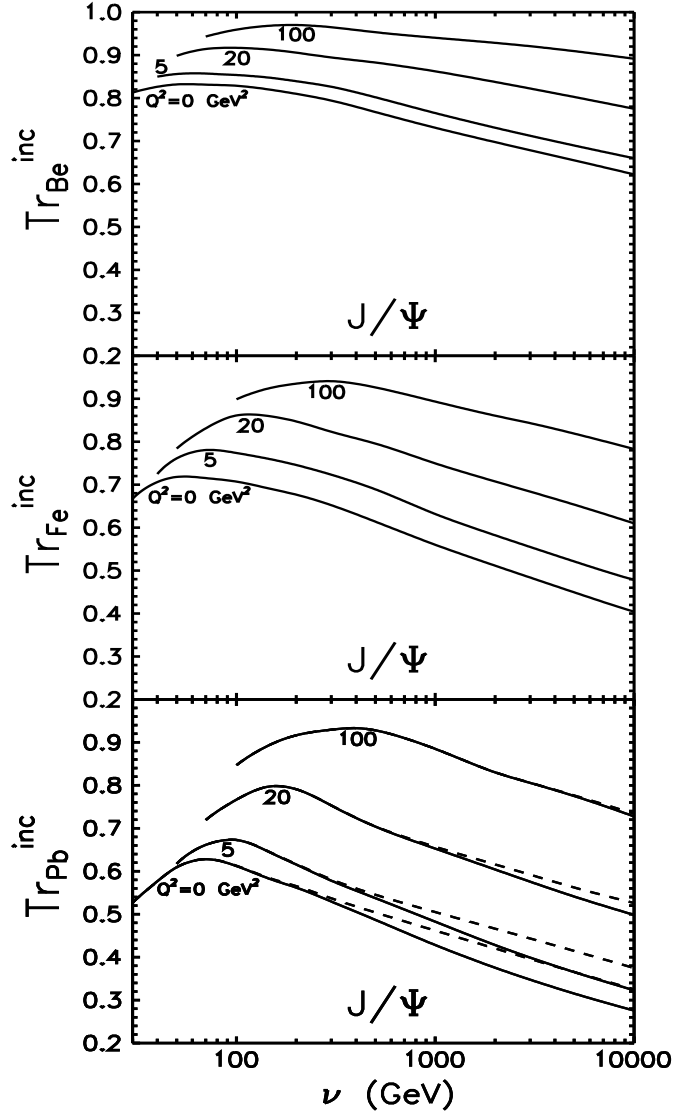}
\begin{center}
\vspace{15.0cm}
\parbox{13cm}
{\caption[Delta]
 {Nuclear transparency for incoherent electroproduction $\gamma^*A\to
J/\Psi~X$ as function of energy at $Q^2=0,\ 5,\ 20$ and $100\GeV^2$ for
beryllium, iron and lead. The solid curves and dashed curves for lead
correspond to calculations with and without gluon shadowing,
respectively.}
 \label{e-incoh}}
\end{center}
 \end{figure}
%

We also calculated the energy dependence of nuclear transparency 
for charmonium production \cite{n-02} at fixed 
$Q^2$ (analogical Fig. for light vector meson production
can be found in \cite{knst-01}).
The results for beryllium, iron and lead are shown  in 
Fig.~\ref{e-incoh} for different values of $Q^2$.
The interesting feature is the presence of a maximum of transparency at
some energy which is much more evident than in production of light
vector mesons \cite{knst-01}.
At small and moderate energies a strong rise of $Tr_{A}^{inc}$ with energy
especially for the lead target 
is a manifestation of net CT effects resulting from a 
strong inequality $l_c < l_f$. 
The existence of maxima of $Tr_{A}^{inc}$
results from the interplay of coherence and formation
effects. Indeed, the formation length  
rises with energy leading to an increasing
nuclear transparency. At some energy, however, the effect of CL 
is switched on leading to a growth of the path
length of the $\bar qq$ in the nucleus, i.e. to a suppression of
nuclear transparency. 
This also explains the unusual ordering of curves 
at small and moderate $Q^2$ calculated for different 
values of $l_c$ as is depicted in Fig.~\ref{lc-const-inc}.

%
\section{Coherent production of vector mesons}
\label{psi-coh}
%

In general,
in coherent (elastic) electroproduction of a vector mesons 
the target nucleus remains intact, so
all the vector mesons produced at
different longitudinal coordinates and impact parameters add up
coherently. This condition considerably simplifies the expressions 
for the production cross sections. The
integrated cross section has the form,
%
 \BE
\sigma_A^{coh}\equiv
\sigma_{\gamma^{*}A\to VA}^{coh} = 
\int d^2q\,\left|\int d^2b\,
e^{i\vec q\cdot\vec b}\,
{\cal M}_{\gamma^{*}A\to VA}^{coh}(b)
\right|^2 = 
\int d^{2}\,{b}\,
|{\cal M}_{\gamma^{*}A\to VA}^{coh}
({b})\,|^{2}\ ,
\label{550}
 \EE
%
 where
%
 \BE
{\cal M}_{\gamma^{*}A\to VA}^{coh}({b}) =
\int\limits_{-\infty}^{\infty}\,dz\,\rho_{A}({b},z)\,
F_{1}({b},z)\ ,
\label{560}
 \EE
%
 with the function $F_{1}({b},z)$ defined in Eq.~(\ref{505}).

One should not use Eq.~(\ref{485}) for nuclear transparency any more
since the $t$-slopes of the differential cross sections for nucleon and
nuclear targets are different and do not cancel in the ratio. Therefore,
the nuclear transparency also includes the slope parameter
$B_{V}$ for the process $\gamma^{*}\,N\rightarrow V\,N$,
%
 \BE
Tr_{A}^{coh} = \frac{\sigma_{A}^{coh}}{A\,\sigma_{N}} = 
\frac{16\,\pi\,B_{V}\,\sigma_{A}^{coh}}{A\,
|{\cal M}_{\gamma^{*}N\to VN}(s,Q^{2})\,|^{2}} \, .
\label{570}
 \EE
%
\vspace*{0.3cm}

%
 \begin{figure}[tbh]
\includegraphics{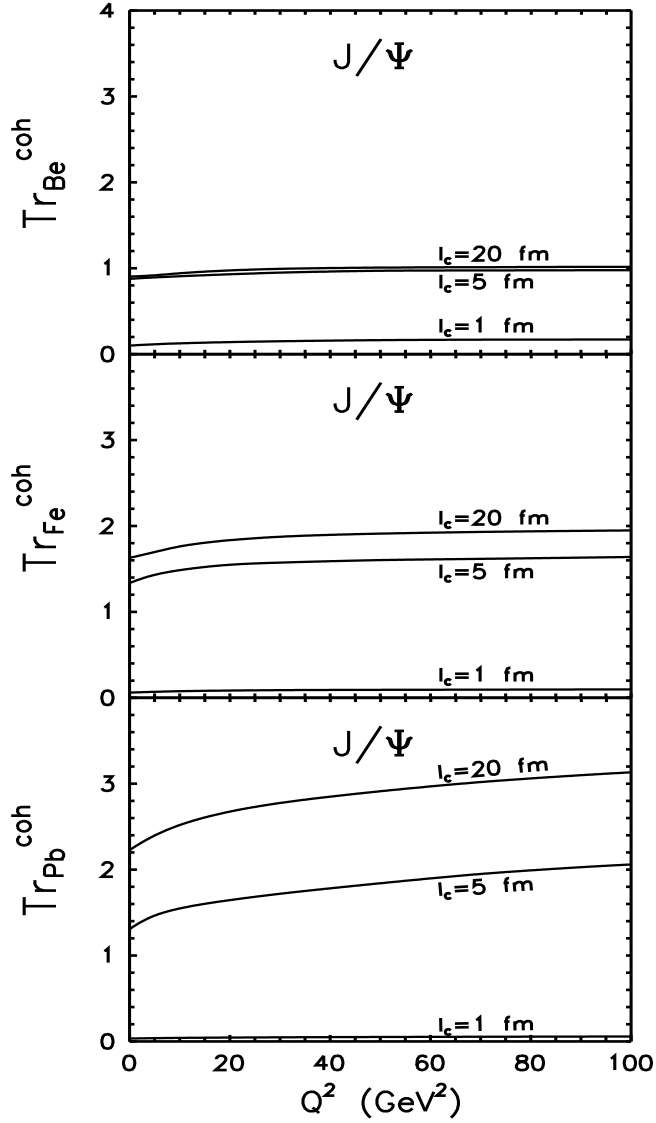}
\begin{center}
\vspace{14.7cm}
\parbox{13.0cm}
{\caption[Delta]
 {The same as in Fig.~\ref{lc-const-inc}, but for coherent production
of $J/\Psi$, $\gamma^*A\to J/\Psi~A$.}
 \label{lc-const-coh}}
\end{center}
 \end{figure}
%

One can eliminate the effects of CL and single out the net CT effect in a
way similar to what was suggested for incoherent reactions by selecting
experimental events with $l_c=const$. We calculated nuclear transparency
for the coherent reaction $\gamma^*A\to J/\Psi A$ at fixed values of
$l_c$. The results for $l_c=1,\ 5$ and $20\fm$ are depicted in
Fig.~\ref{lc-const-coh} for several nuclei.
We performed calculations of $Tr_A^{coh}$ with the slope 
$B_V = B_{J/\Psi} = 4.7\GeV^{-2}$.
The effect of a rise of $Tr_{A}^{coh}$ is not sufficiently large to be 
observable in the range of $Q^2\leq 20\GeV^2$.
A wider range of $Q^2\leq 100\GeV^2$ and heavy nuclei gives
higher chances for experimental investigation of CT. However,
on the other hand, it encounters the problem of low yields at high $Q^2$.

%
 \begin{figure}[htb]
\includegraphics{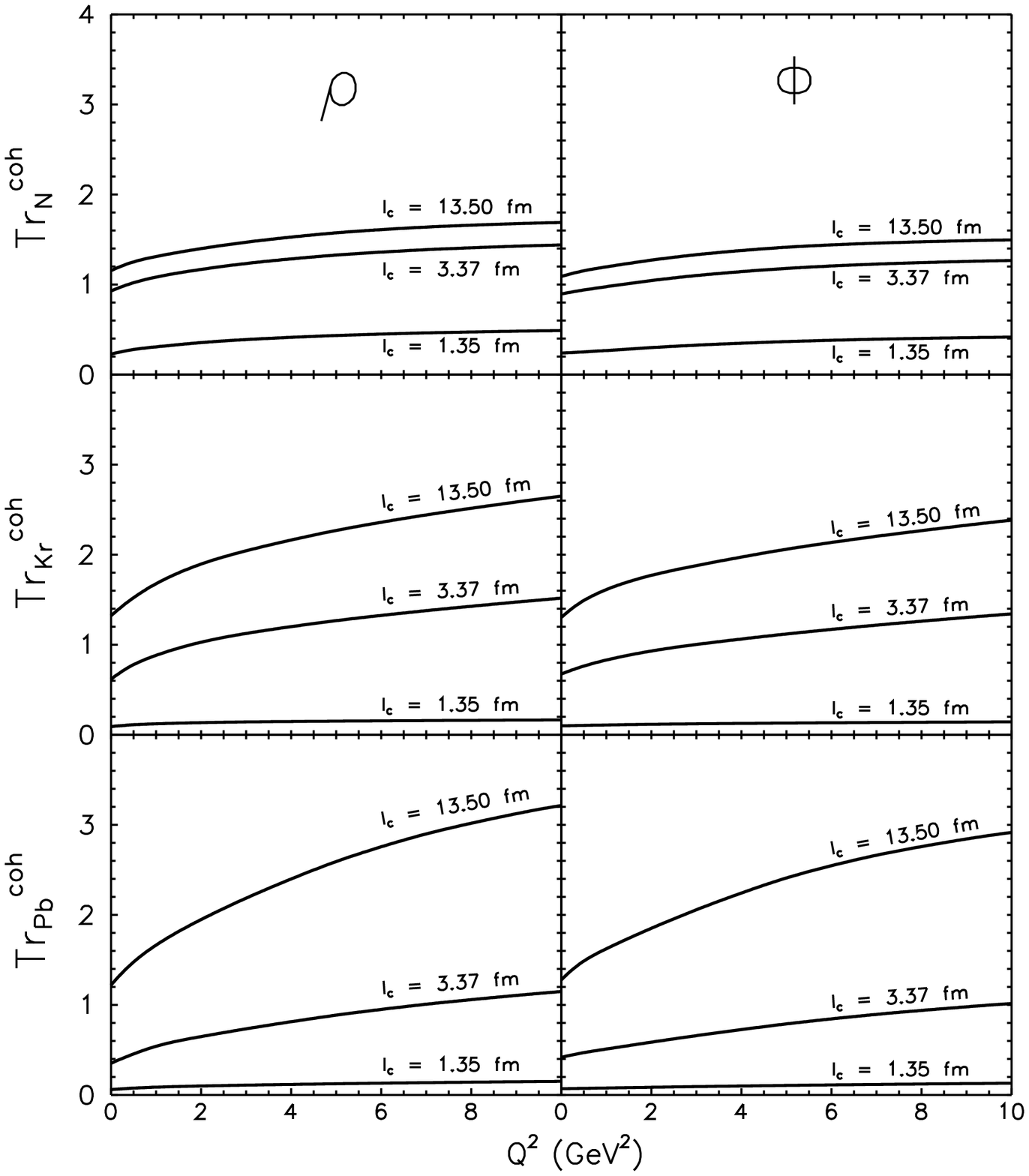}
\begin{center}
\vspace{15.5cm}
\parbox{13cm}
{\caption[Delta]
 {The same as in Fig.~\ref{lc-const-inc-all}, but for
coherent production of $\rho$ and $\Phi$,   
$\gamma^*A\to V~A$.}
 \label{lc-const-coh-all}}
\end{center}
 \end{figure}
%

We calculated also nuclear transparency for coherent production
of light vector mesons at fixed values of $l_c$. 
We took parametrization of the slope parameter from \cite{knst-01}.
The results
for $l_c=1.35, 3.37$ and $13.50\fm$ are presented in 
Fig.~\ref{lc-const-coh-all} \cite{knst-01}.

Note that in contrast to incoherent production where nuclear transparency
is expected to saturate as $Tr^{inc}_A(Q^2) \to 1$ at large $Q^2$, for
the coherent process nuclear transparency reaches a higher limit,
$Tr^{coh}_A(Q^2) \to A^{1/3}$.

%
 \begin{figure}[htb]
\includegraphics{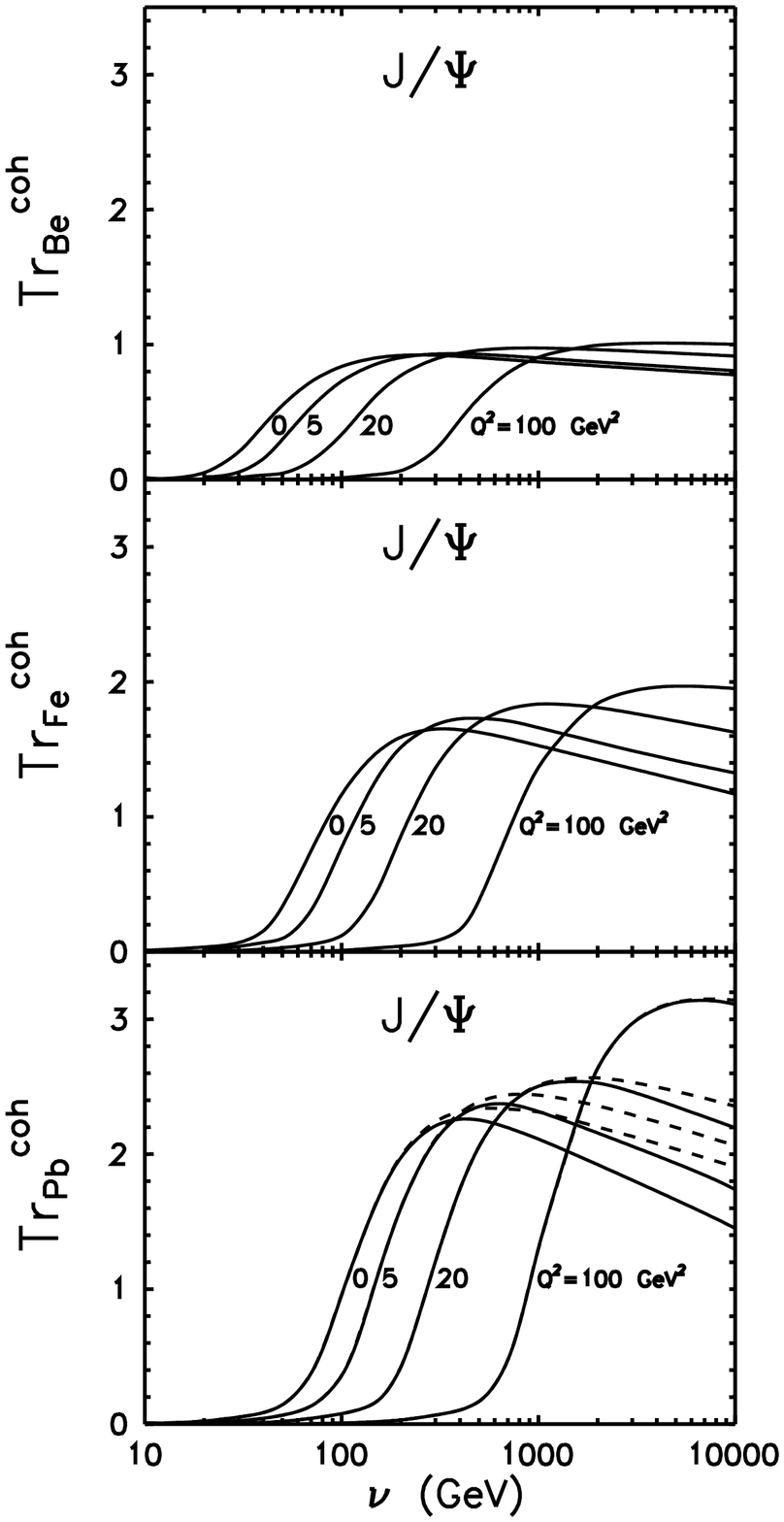}
\begin{center}
\vspace{14.7cm}
\parbox{13cm}
{\caption[Delta]
 {Nuclear transparency for coherent electroproduction   
$\gamma^*A\to J/\Psi~A$ as function of energy at $Q^2=0,\ 5,\
20$ and $100\GeV^2$
for beryllium, iron and lead.
The solid curves and dashed curves for lead
correspond to calculations with and without gluon shadowing, 
respectively.}
 \label{e-coh}}
\end{center}
 \end{figure}
%

We also  calculated nuclear transparency as function of energy at fixed 
$Q^2$.
The results for $J/\Psi$ produced coherently off beryllium, iron
and lead are depicted 
in Fig.~\ref{e-coh} at $Q^2=0,\ 5,\ 20$ and $100\GeV^2$
(analogical Fig. for coherent production of light vector mesons
can be found in \cite{knst-01}). 
At low energy, $Tr^{coh}_A$ is very small due to suppression of 
the nuclear coherent cross section
by the nuclear form factor. 
The reason is that 
the longitudinal momentum transfer, which is equal to the 
inverse CL, is large when the CL is short. 
At high energy, however, $l_c\gg R_A$ and nuclear
transparency nearly saturates (it decreases with $\nu$ only due to the
rising dipole cross section). The saturation level is higher at larger
$Q^2$ which is a manifestation of CT.

Note that in all calculations the effects of gluon shadowing are 
included in a way analogical to that   
described in the recent papers
\cite{knst-01,ikth-02}.
For illustration they are depicted in Figs.~\ref{e-incoh} and
\ref{e-coh} for the lead target 
as a difference between solid and dashed lines 
at various values of $Q^2$.
For charmonium production, 
in the photoproduction limit $Q^2 = 0$ 
the onset of gluon shadowing becomes
important at rather high photon energy $\nu > 1000\GeV$
for incoherent and $\nu > 500\,\GeV$ for coherent
production. In the production of light vector
mesons \cite{knst-01}, however, the onset of gluon shadowing
is important at much smaller energies due to
much smaller masses of $\bar qq$ fluctuactions of
the photon.
This corresponds to the
claim made in \cite{kst2} that the onset of gluon shadowing requires
smaller $x_{Bj}$ than the onset of quark shadowing. The reason is 
that the fluctuations containing gluons 
are in general heavier than the $\bar qq$
and have a shorter CL.

%
\section{Summary and conclusions}
\label{conclusions}
%

We presented a rigorous quantum-mechanical approach based
on the light-cone QCD Green function formalism
which naturally incorporates the interference 
effects of CT and CL.
Within this approach \cite{n-02,knst-01} 
we studied CT effects in coherent and incoherent 
electroproduction of light vector mesons and charmonia off nuclei.

The onset of coherence effects (shadowing) can mimic the expected
signal of CT in incoherent electroproduction of vector mesons at medium 
and large energies. In order to single out the formation effect the data must
be taken at such energy and $Q^2$ which keep $l_c = const$. 
Then the observation of a rise with $Q^2$ of nuclear
transparency for fixed $l_c$ would give a signal of color
transparency.
Predictions of $Tr_A^{inc}(Q^2)$ as a function of $Q^2$
at different fixed $l_c$ show rather large CT effects
in incoherent production of light vector mesons and charmonia.
Although for charmonium production \cite{n-02} 
the $Q^2$- variation of nuclear transparency 
at fixed $l_c$ is predicted to be less 
than for the production of light vector mesons 
\cite{knst-01}, it is still sufficiently significant
to be investigated experimentally even in the range of $Q^2\lsim
10\div 20\GeV^2$.
CT effects (the rise with $Q^2$ of nuclear transparency)
are more pronounced at low than at high energies
and can be easily identified by HERMES and planned future
experiments.  

The effects of CT in coherent production of vector mesons are found
to be less pronounced.
A wider range $Q^2\leq 100\GeV^2$ and heavy nuclei give
higher chances for experimental investigation of CT. However,
on the other hand, it faces the problem of low yields at high $Q^2$.

Nuclear suppression of gluons was calculated
within the same LC approach and included in predictions.
It was manifested that these corrections are quite small 
at medium energies which are dominant
in the process of searching for CT effects.

Concluding, the predicted large effects of CT in 
electroproduction of light vector mesons and charmonia off nuclei 
open further 
possibilities 
to search for CT with medium energy electrons and can be tested
at HERMES and in future experiments.

\medskip

\noindent
 {\bf Acknowledgments}:
This work has
been supported in part by the Slovak Funding Agency, Grant No. 2/2099/22.

\end{document}